\documentclass{ws-ijmpa}
\usepackage[super,compress]{cite}
\usepackage{graphicx}
\usepackage[utf8]{inputenc}
\usepackage{hyperref}
\usepackage{xcolor}
\usepackage{lipsum}
\usepackage[normalem]{ulem}

\newcounter{addref}

\newcounter{FPcomment}
\newcommand{\FP}[2][]{%
\refstepcounter{FPcomment}{%
\todo[color={red!60},size=\small]{%
\textbf{FP [\uppercase{#1}\theFPcomment]:}~#2} }}

\newcounter{MGcomment}
\newcommand{\MG}[2][]{%
\refstepcounter{MGcomment}{%
\todo[color={red!40!blue!20},size=\small]{%
\textbf{MG [\uppercase{#1}\theMGcomment]:}~#2} }}

\newcounter{KWcomment}
\newcommand{\KW}[2][]{%
\refstepcounter{KWcomment}{%
\todo[color={blue!40},size=\small]{%
\textbf{KW [\uppercase{#1}\theKWcomment]:}~#2} }}

\newcounter{CTcomment}

\begin{document}



\markboth{F. Psihas, M. Groh, C. Tunnell, K. Warburton}{Machine Learning and Neutrinos}

%
\catchline{}{}{}{}{}
%
\title{ A Review on Machine Learning for Neutrino Experiments}

\author{Fernanda Psihas}

\address{Neutrino Division, Fermi National Accelerator Laboratory\\
Batavia, Illinois, United States of America\\
psihas@fnal.gov}

\author{Micah Groh}

\address{Department of Physics, Indiana University\\
Bloomington, IN, United States of America\\
mcgroh@iu.edu}

\author{Christopher Tunnell}

\address{Department of Physics and Astronomy, RICE University\\
Houston, Texas, United States of America\\
tunnell@rice.edu}

\author{Karl Warburton}

\address{Department of Physics and Astronomy, Iowa State University\\
Ames, Iowa, United States of America\\
karlwarb@iastate.edu}

\maketitle

\begin{history}
Submitted to IJMPA 3 August 2020
\end{history}

\begin{abstract}
Neutrino experiments study the least understood of the Standard Model particles by observing their direct interactions with matter or searching for ultra-rare signals. 
The study of neutrinos typically requires overcoming large backgrounds, elusive signals, and small statistics. 
The introduction of state-of-the-art machine learning tools to solve analysis tasks has made major impacts to these challenges in neutrino  experiments across the board. 
Machine learning algorithms have become an integral tool of neutrino physics, and their development is of great importance to the capabilities of next generation experiments. 
An understanding of the roadblocks, both human and computational, and the challenges that still exist in the application of these techniques is critical to their proper and beneficial utilization for physics applications.
This review presents the current status of machine learning applications for neutrino physics in terms of the challenges and opportunities that are at the intersection between these two fields.

\keywords{Neutrinos; Machine Learning; Deep Learning; Review.}
\end{abstract}

\ccode{PACS numbers:}

\newpage

\section{Introduction} 

The nature of neutrinos and their masses is one of the main science drivers of particle physics today~\cite{P5:2014,APPEC:2017}.
Not only are neutrinos the least understood particle in the Standard Model, they may be linked to the explanation of the matter/antimatter asymmetry in the Universe through the process of leptogenesis~\cite{Pascoli:2006ie}. 
Neutrinos exhibit unexpected oscillations between their mass states, a behavior which indicates that other new physical phenomena beyond the Standard Model might be possible.
Specifically, it raises the question about the mechanism through which they acquire the non-zero mass required by oscillations as well as the possibility that neutrinos engage in charge-parity (CP) violating processes, both linked directly to leptogenesis. 
The answers to questions about the mass mechanism and CP violation can provide a deeper understanding of the early Universe through the study of neutrinos.

In the aftermath of the solar neutrino problem~\cite{solarnu}, resolved by the discovery of oscillations~\cite{oscdiscovery,sno}, a large number of experiments have set out to answer the remaining questions of neutrino physics, taking advantage of the best particle detection technology available to them. 
The low cross-sections of neutrino interactions and the background suppression required by many of these experiments makes the study of neutrinos technically challenging and subject to statistical limitations. 
The optimization of signal and background separation, detection threshold, and physics reconstruction, are all key factors in the technology design for a particular experiment. 

The software tools used for analysis and reconstruction of detector data are often overlooked as key components of experimental technology.
These tools are not only used for analysis and final results, but also form an integral part of the conception and design of new projects. 
Improvements in reconstruction and analysis technologies have enhanced our ability to extract information from data and translate it into physics quantities.
The study and development of reconstruction and analysis tools is, thus, of critical importance to the capabilities of particle physics experiments.
The tools of machine learning, also broadly referred to as artificial intelligence, have been at the center of analysis techniques for several decades.


Beyond appealing to the personal interest of the reader in machine learning, it is clear from the abundance of applications of these tools, that a basic knowledge of machine learning is central to the understanding of experimental data analysis in neutrino physics today.
In this manuscript, we review the algorithms, developments, and evolution of machine learning tools for neutrino experiments with a focus on deep learning. 
We discuss the obstacles that still challenge the standing of these tools as trusted parts of the experimentalist's arsenal. 
We first define some essential language and formalism for this discussion and introduce the basic components of machine learning algorithms and concepts of deep learning. 
The current status and prospects in the field will be discussed in terms of the  roadblocks, both human and computational, and the challenges and opportunities that still exist in the application of these techniques.


\section{Machine Learning and Deep Learning}
\label{sec:pre-ml}

The term machine learning is an umbrella term for all algorithms where inference is used to perform a task and that have the ability to improve with experience, though terms like deep learning are commonly used depending on the complexity of these algorithms. 
Within the past decade, deep learning algorithms have gained significant popularity in neutrino experiments and have enabled large improvements in the performance and physics reach of the analyses where they are employed.

A common entry-point to the understanding of machine learning algorithms is the description of an artificial neural network or ANN. 
ANNs are interconnected series of elementary functions called neurons, somewhat analogous to the biological system of the brain. 
Artificial neurons employ mathematical functions called activation functions to produce an output, mimicking the action potential which determines the production of electrical signals in brain cells. 
The connections between artificial neurons, where the output of each one is passed as input to others, allow the network to combine these simple units to perform the complex task of learning.
Similarly, connecting a large number of artificial neurons results in interesting macroscopic behavior, as discussed in the following sections.

\begin{figure}
    \centering
    \includegraphics[width=0.55\textwidth]{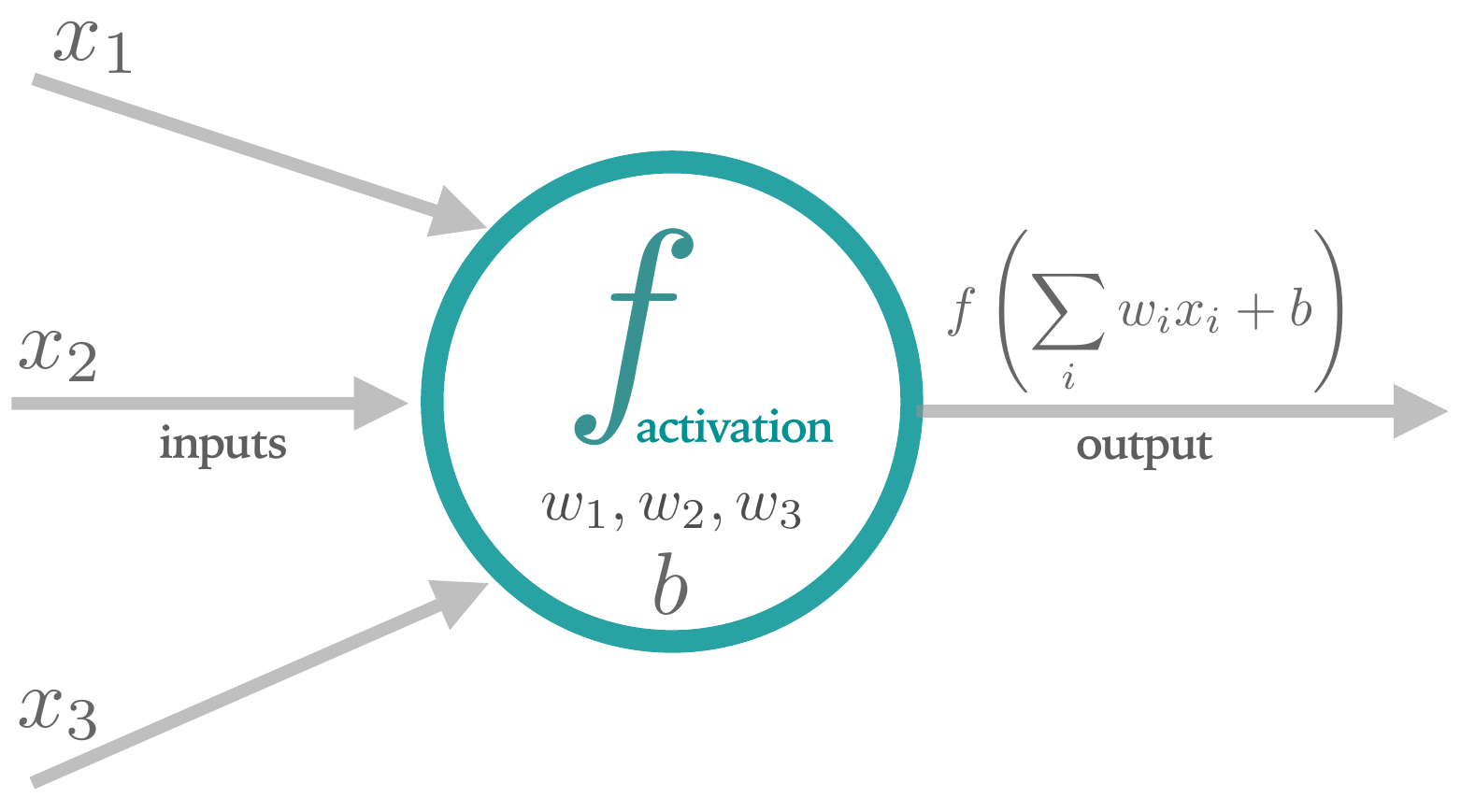}
    \includegraphics[width=0.44\textwidth]{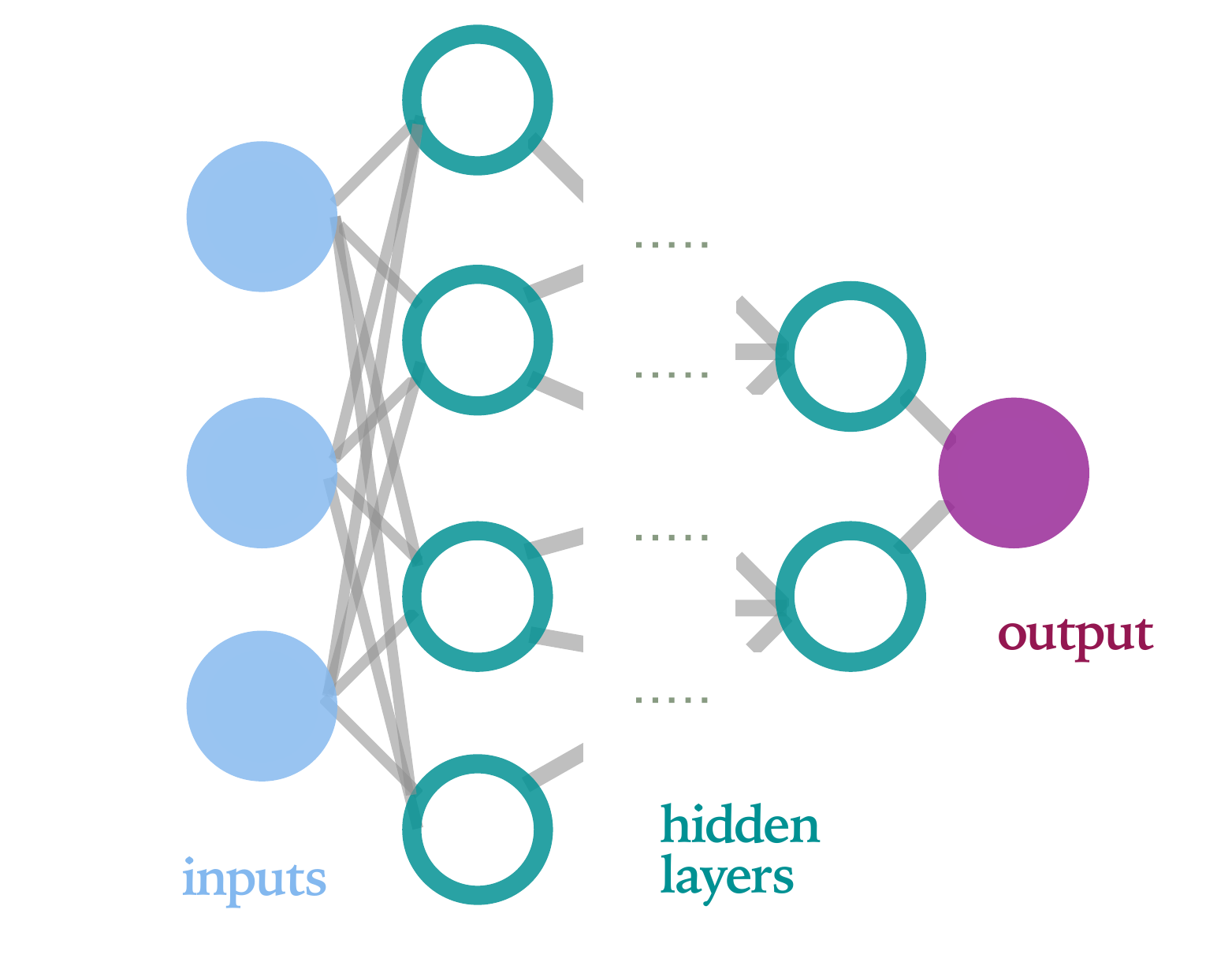}
    \caption{Left: The structure of a neuron, the building block of artificial neural networks. Each neuron takes as input the values output from the previous layer. The values are combined in a linear a combination, some bias is applied, and then a non-linear activation function is applied. Right: A complete artificial neural network. Each neuron is connected to every neuron in the previous layer and every neuron in the final layer.}
    \label{fig:nn}
\end{figure}

In the mathematical representation, each neuron receives one or more inputs which are individually weighted. The output is determined by the activation function, which is a nonlinear transformation of the inputs.
In ANNs, neurons are placed into connected layers, often as seen in Figure~\ref{fig:nn}. 
The Multi-layer perceptron MLP depicted in the figure is a type of ANN in which the neurons are organized into ``hidden layers" between the input and the output. In this type of network, also called feedforward network, the  outputs of each layer are fed to the next layer, and so on. 
The number of hidden layers and their interconnected array of neurons allows the network to perform complex tasks. 
For example, a network can be used to reproduce a mapping $F$ of an input vector $\vec{x}$ to an output vector $\vec{y}$.
The input data used by the network to learn $F$ are a collection of ``ground truth'' examples for which the input $\vec{x}$ and the exact output $\vec{y} = F(\vec{x})$ are known or have been simulated.

The process of learning occurs iteratively by first constructing output estimates for given values of $\vec{x}$ using a set of initial weights for each component of the network.
Then, the difference between these estimates $F'\left(\vec{x}\right)$ computed by the network and the desired target are minimized by introducing changes to the weights of each neuron.
How the differences, or losses, are quantified and the choice of minimization function vary by application.
The iterations are repeated until the ability of the network to approximate the function's behavior no longer improves.

Note that while the task of the network is to reproduce the output of the target function given the same set of inputs, it need not know or approximate the exact form of the function to accomplish that task. 
For example, a network trained to reproduce the invariant mass of an initial state does not need to know or learn decay kinematics, but instead it learns to reproduce the same principles by training on final state vectors with corresponding invariant mass values.
Therefore, it is possible for these networks to perform complex tasks through the same simple learning process, which with several computational innovations have become a powerful tool for a variety of tasks.

Other common tasks performed by these algorithms on detector data traditionally include regression and classification.
Regression involves learning the mapping between dependent and independent variables, where this is often a continuous mapping predicting quantities such as particle energy. 
Classification involves learning the category associated with the data, which in common applications is used for signal and background discrimination. 
Classification is done by normalizing scores for each category to sum to unity using a softmax activation function. 
See \refcite{nwankpa2018activation} for a description of many common activation functions used in neural networks.
The tasks performed by these algorithms are more complex and increasingly take on more of the process of reconstruction and analysis of detector data.

\subsection{Deep Learning}
\label{sec:new-ml}

Deep learning algorithms is differentiated from machine learning by the complexity of the algorithms used.
Deeper network induce more non-linear operations such that the mapping from results to the input variables is more challenging to track.
Deep learning algorithms have gained popularity in the last two decades due to breakthroughs in their performance, largely enabled by the rapid development of hardware such as graphics processing units (GPUs). 
The field of computer vision has been the primary driver of the innovations, which are used to solve pattern recognition tasks~\cite{visunderstand}. 

Deep neural networks are sophisticated and more computationally expensive techniques which are able to tackle problems of higher complexity than other machine learning tools. 
Increasing network depth by adding additional layers, for instance, allows for the approximation of increasingly complicated functions.

One of the most common deep learning algorithms employed for pattern recognition is the Convolutional Neural Network (CNN)~\cite{Schmidhuber_2015}. 
CNNs are a class of MLP which learn to extract features from an input in addition to training for the intended task. 
These features are identified using the spatial relationship between neighboring regions in the image. 
The key components of CNNs are kernels, or image filters~\cite{oshea2015introduction}, which are matrices that scan an input image and output an image with highlighted features. 
A convolution layer consists of operating one or more kernels across an input image. 

\begin{figure}
    \centering
    \includegraphics[width=0.99\textwidth]{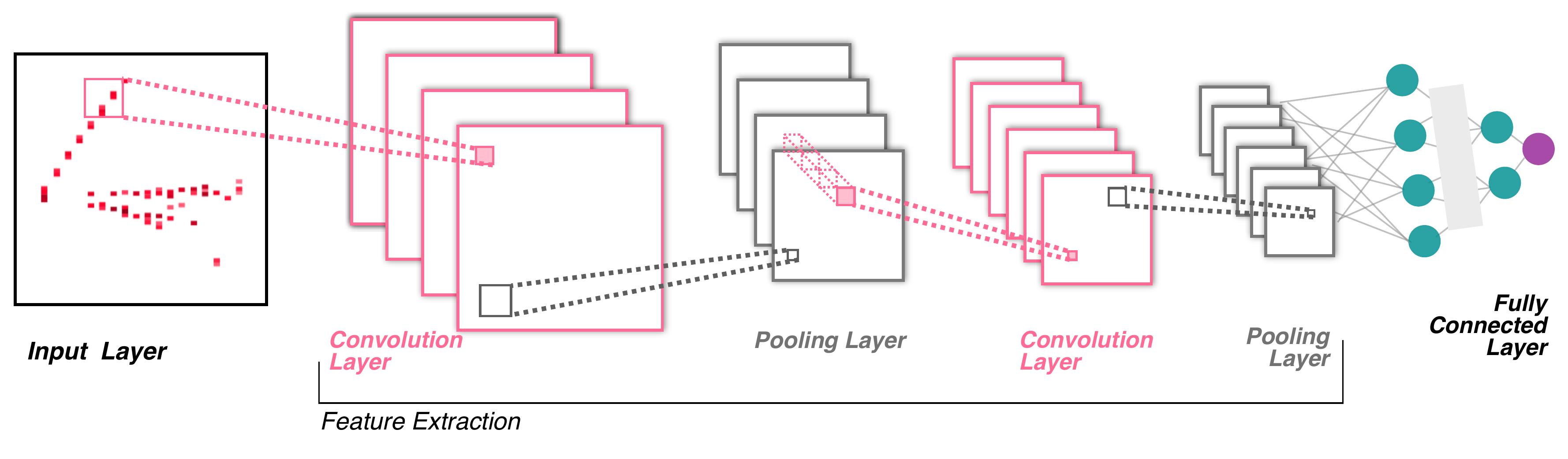}
    \caption{The structure of a convolutional neural network. The convolution layers use image kernels to extract features from the input. The pooling layers downsample the image. The final set of features are connected to a fully connected, artificial neural network.}
    \label{fig:cnn}
\end{figure}

In reality, the input to CNNs are tensors typically containing the  pixel-by-pixel RGB values of an image. 
The dimensions and content of the input tensors can be altered for different applications but most developments in image recognition naturally use the image-to-RGB tensor strategy. 
Because convolutions output the effect of a kernel on the input tensor with translational invariance, they are especially useful for image and pattern recognition, where the features of interest are topological characteristics. 



 
Figure~\ref{fig:cnn} shows the basic structure of a CNN. 
As seen in the figure, the fully connected layers of a CNN are notably similar to the basic MLP, whereas the initial convolutional layers serve the purpose of feature extraction. 
The kernel values are learned during the training process to extract features that are most useful for the desired task.
Convolution layers are often interlaced with pooling layers which downsample the image to reduce the computations needed deeper in the network and promote translational or rotational invariance.
The final layers of the network then perform the classification or regression task using the extracted features as input.



 
The task of identifying signals and reconstructing physical characteristics of interactions in particle detectors is often analogous to that of pattern recognition in images. 
Thus, much of the recent development in applications in neutrino physics involves usage or adaptations of deep learning networks developed for image recognition.  
Within the past decade, deep learning algorithms have gained significant popularity in neutrino experiments and have enabled drastic improvements in the performance and sensitivity of the analyses where they have been employed.  
Deep CNNs have now demonstrated state of the art performance on many tasks~\cite{alom2018history} and are one of the most common tools used in neutrino physics.
The advantages and motivation to use  CNNs in neutrino experiments are largely applicable to other deep neural networks used in neutrino experiment as well. Similarly, the  advantages, and the challenges discussed in the following section applies to CNNs. 

\section{Applications in Neutrino Experiments}
\label{sec:rise-ml}

\begin{figure}
    \centering
    \includegraphics[width=0.8\textwidth]{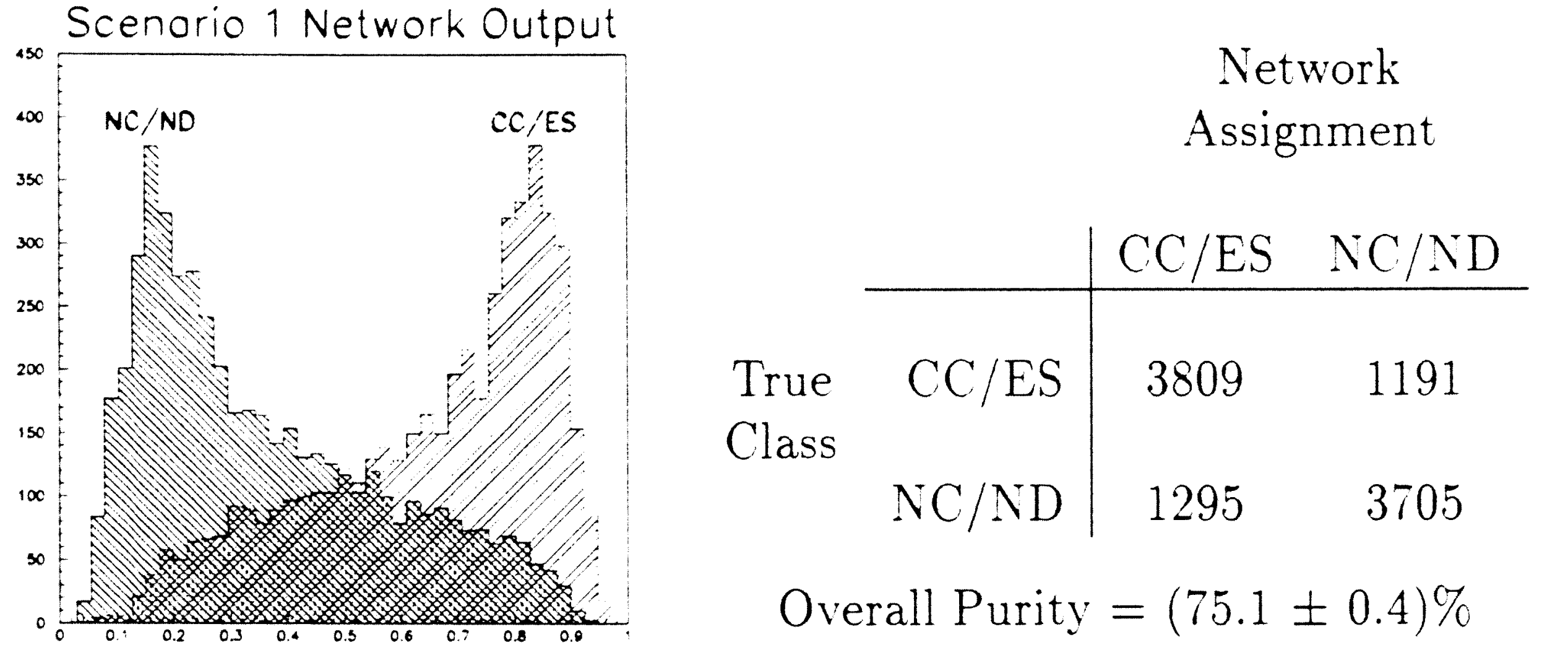}
    \caption{First demonstration of neural networks for neutrino physics. The network was trained to separate charged current and neutral current neutrino interactions for the SNO experiment. The table on the performance of the network with a classification matrix.}
    \label{fig:sno}
\end{figure}

Particle physics experiments, including neutrino experiments, are endeavours which require the analysis of large data sets, sophisticated modeling, and statistics.
In the past two decades, both neutrino physics and machine learning have been experiencing a renaissance with the discovery of neutrino oscillations and the advent of deep learning, respectively.
In the 1990s the initial exploration of neural networks in particle physics began~\cite{1990PhRvL..65.1321L}.
The SNO experiment was the first to explore the use of neural networks in neutrino physics~\cite{SNO_STR_96}, using feedforward networks, a type of artifical neural network to classify events based on hit pattern features, shown in Figure~\ref{fig:sno}.

While it is true that these neural networks did not outperform other statistical techniques at first, they demonstrated the capabilities of these techniques for event classification in neutrino detector data.
As expertise grew regarding the impact of sample preparation and feature choices in network performance, not only did machine learning techniques surpass traditional reconstruction, but they would grow to be one of the most widely used analysis techniques in the field. 

Machine learning has played a role in nearly every particle physics discovery and measurement since.
Common analysis frameworks designed for particle physics have natively supported the use of these tools for almost two decades~\cite{Brun:1997pa}.

The role of machine learning in physics analyses has only grown in scope, taking advantage of several opportunities specific to our problem set which will be discussed in the next section.
These first applications are now commonplace in our field, typically using tools like feedforward networks, MLPs, and more recently boosted decision trees, to name a few. 
Most common applications start with input variables which have been pre-extracted and selected by the analyzer. 
This continued as the main strategy until the introduction of deep learning tools.




The tasks that deep learning algorithms have been applied to in the last decade span the full extent of experimental analysis work flow, including design, hardware triggers, energy estimation, reconstruction, and signal selection. 
Many applications exist which have greatly simplified and improved the performance of experiments and their physics reach when compared to the standard tools they have replaced. 
The performance achieved by these tools is the prime motivation for their implementation to solve physics problems, despite the computational complications which will be described later in this section.

Even more significant than the improvements themselves are the implications of the usage of these tools in our experiments. 
The interplay between neutrino physics and deep learning is rich in both challenges and opportunities for both fields. 
The current status of the field is presented in this section, in the context of these challenges and opportunities. 
Rather than providing an exhaustive list of applications in a rapidly growing field, those that are notable are highlighted when relevant to the item discussed. 

\subsection*{Challenge 1 --- Adaptability of the Methods}

The most frequently used deep learning algorithms in neutrino experiments are those developed or commonly used for image recognition. 
Given that some experimental setups closely resemble or can be mapped into 2-dimensional images, this is a natural starting point for many studies to apply the tools of image recognition. 

However analogous, the problems solved for image or pattern recognition have important differences with particle physics.
Some adaptation is usually required for the usage of these algorithms. 
Adaptation can be as simple as converting detector data into image-like tensor inputs or as complicated as complete network redesign for the new task. 
The trade-off between simple adaptation and those where the inputs and network are more tuned to the particular task can be significant in terms of performance improvement.

An example of a deep learning network used with different adaptations in neutrino experiments is the GoogLeNet~\cite{szegedy2014going} CNN architecture.  
GoogLeNet was the first creatively non-sequential implementation of convolutional layers in CNNs, which brought significant accuracy and performance improvements with respect to its competitors~\cite{ILSVRC15}. 
Following the success of GoogLeNet, many neutrino experiments explored its utilization with minimal or no modifications as a starting point for their own classification studies. 
Despite the differences between images and neutrino data, out-of-the-box approaches yielded important successes over traditional methods. 

\begin{figure}
    \centering
    \includegraphics[width=0.9\textwidth]{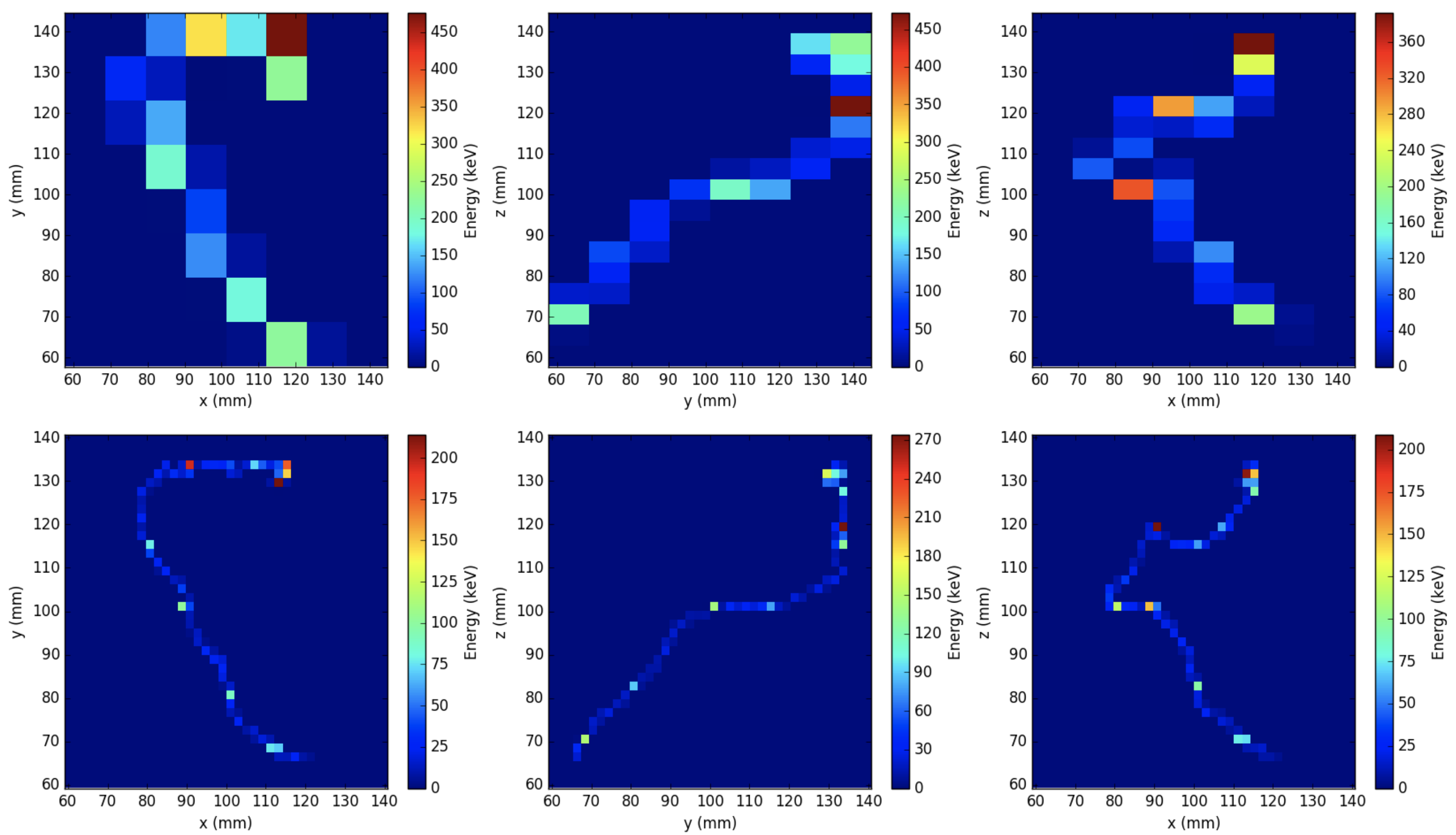}
    \caption{Input data for the NEXT CNN classifier. Top: Example event in NEXT  with 10 mm voxelization. Bottom: Example event in NEXT with 2 mm voxelization. Columns are the $xy$, $yz$, and $xz$ views of the event. The distinguishing feature of the track projection that identifies these as signal events is the presence of a larger energy deposition (bragg peak) at the end of each track. This feature is mostly lost in the 10 mm voxelization.}
    \label{fig:next}
\end{figure}

A successful out-of-the-box application of GoogLeNet is the NEXT experiment background rejection network~\cite{Renner_2017}. 
The NEXT detectors are cylindrical time projection chambers with photon detection and charge detection at each end, respectively. 
Photomultiplier tubes collect a light signal and silicon photomultipliers (SiPM) collect an electroluminescence signal from drifted charges inside the detector~\cite{Alvarez:2012sma}.
For the training inputs to resemble 2D images of the particle tracks, the granularity of the data from the SiPM readout is reduced to 3D voxels, of dimensions $x$,$y$ (spatial) and $z$ (drift time), which are used as the RGB channels of the CNN input tensor. 
The inputs for different voxel sizes are shown in Figure~\ref{fig:next}.
Equal numbers of simulated neutrino-less double beta decay signal and radioactive background events are used for the training. 
This simple implementation was found to outperform the traditional reconstruction by between a factor of 1.2 and 1.6 depending on the reconstruction resolution. 

The many differences between 2D images and detector data provide an opportunity to improve algorithm performance by making thoughtful modifications to the original networks. 
In many cases, large improvements have been attained from enhancing useful features of the data by making changes to the algorithms and the structure of the inputs. 

\begin{figure}
    \centering
    \includegraphics[width=0.6\textwidth]{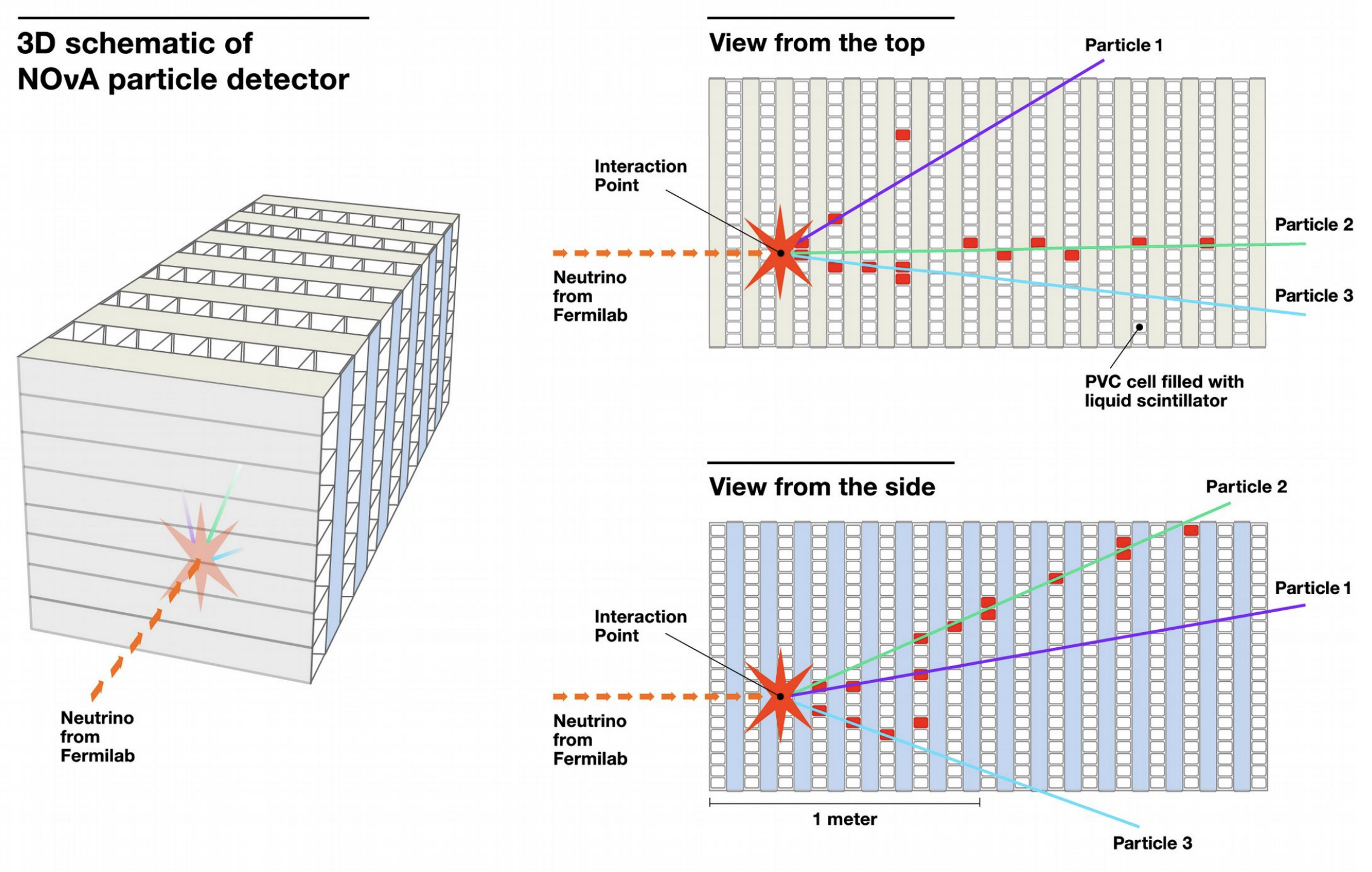}
    \includegraphics[width=0.35\textwidth]{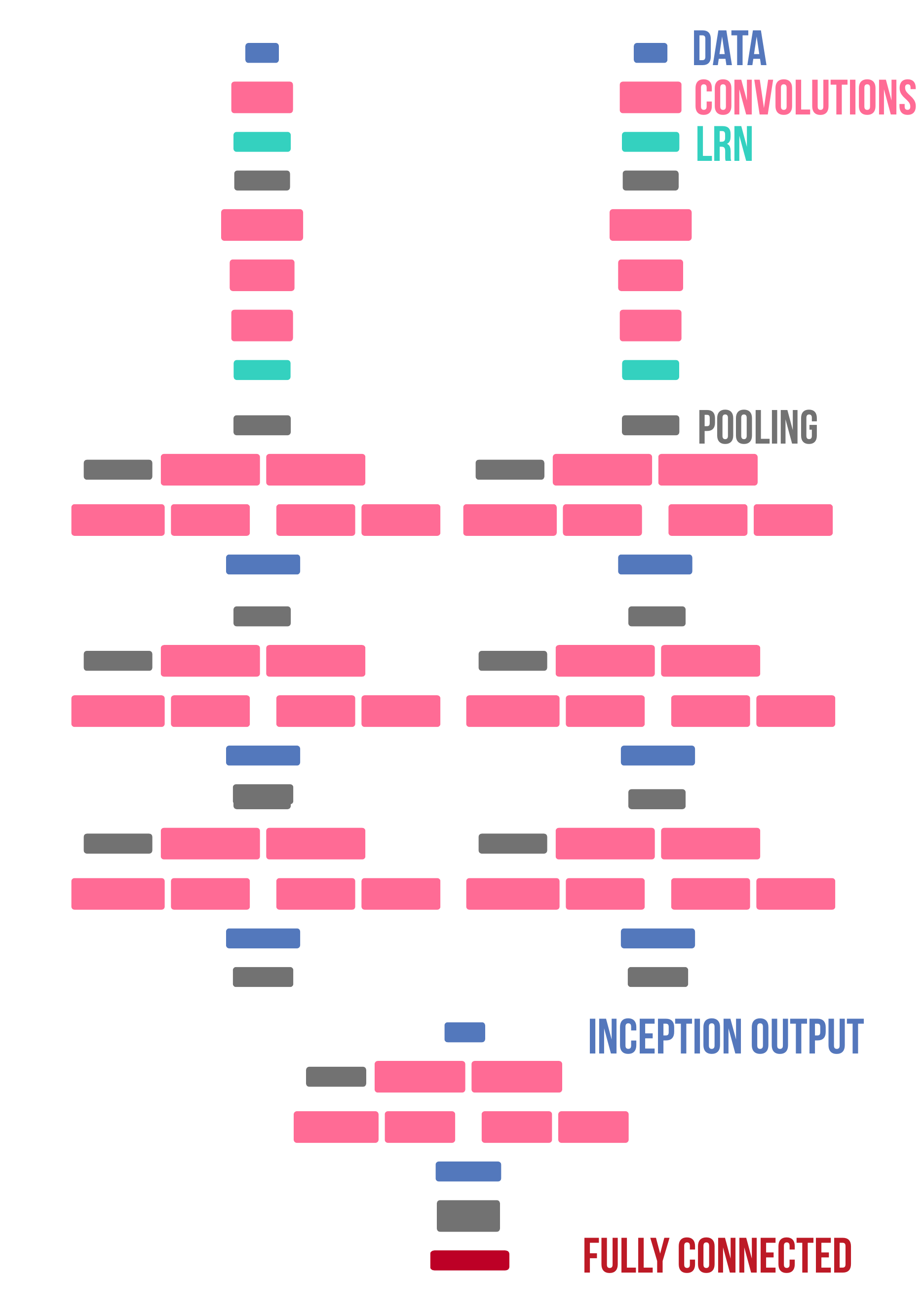}
    \caption{Left: An example readout from the NOvA detector. The planes are arranged in alternating orientations to give two orthogonal views of the event. Right: Siamese tower structure of NOvA's Convolutional Visual Network, based on GoogLeNet, for neutrino flavor classification. The two towers independently operate on each view of the event. The features from each tower are concatenated in the final layers of the network.}
    \label{fig:nova}
\end{figure}

Such is the case for the Convolutional Visual Network~\cite{Aurisano_2016}, a CNN classifier designed for application on NOvA data. 
The readout from the two orthogonal views of the NOvA detectors is already very image-like and naturally depicts 2-dimensional projections of energy depositions. 
However, the decoupled nature of the two views makes a simple conversion to a single RGB tensor unideal because a conversion of the $xz$ and $yz$ views into RGB channels of the same image tensor would result in an unnatural overlap of unrelated features.
Rather than artificially overlapping the orthogonal views, the authors employed a Siamese network structure\cite{siameseifyouplease}, allowing independence in the learning from each detector view to identify neutrino interaction flavor. 
Figure~\ref{fig:nova} shows NOvA's detection technology and the architecture used for neutrino identification.
This, among other modifications to a GoogLeNet-inspired architecture produced large accuracy improvements, increasing the effective exposure of the experiment by 30\%.
This network was the first to be used in a published physics result~\cite{Adamson:2017gxd}, and it demonstrated the significance and impact of adapting both the tool and the inputs to the detector technology.

In addition to detector technology and readout, the geometry of the detector is also relevant to the adaptability challenge.
In some cases, thoughtfully considering modifications based on detector geometry can boost performance significantly.
In other cases, this consideration could be essential for applying the tools with any success.
Additionally, careful consideration in how to map detector readout to inputs compatible with the network of choice should not me overlooked.

\begin{figure}
    \centering
    \includegraphics[width=0.8\textwidth]{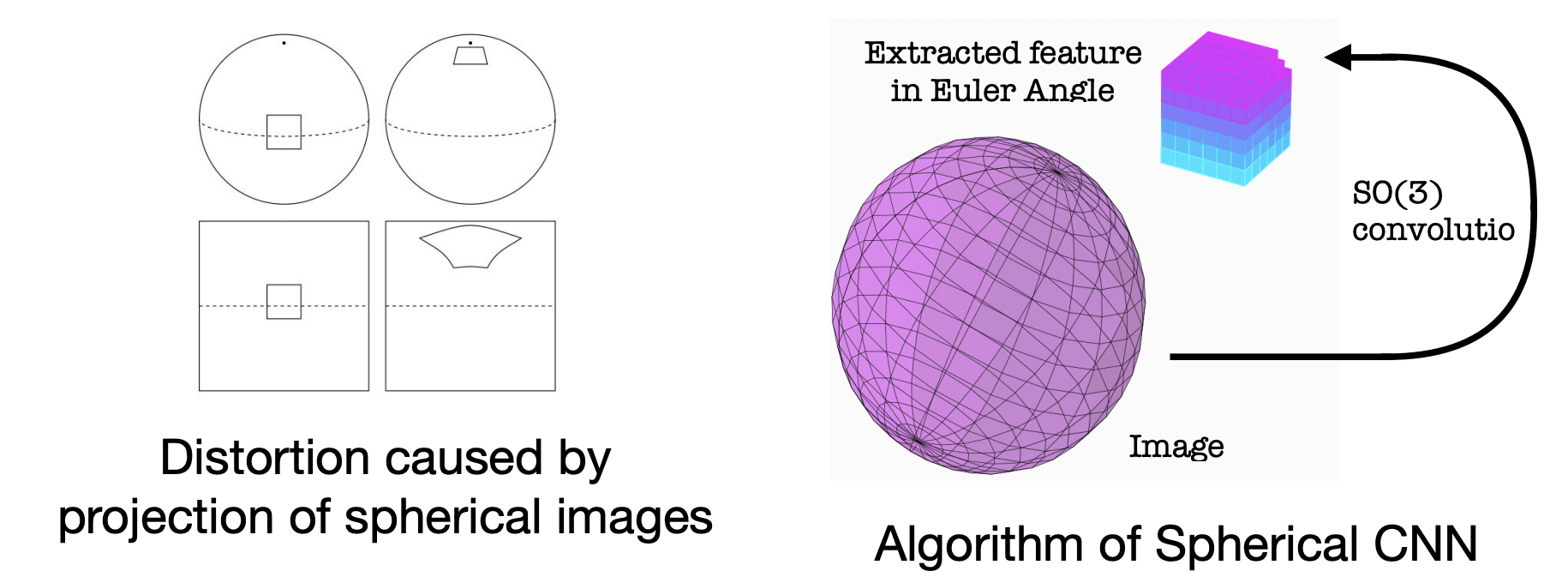}
    \caption{The Kamland-Zen experiment uses spherical convolutions for signal and background separation. The detector is nearly spherical so a traditional mapping to 2D would cause abnormal distortions in the data.}
    \label{fig:kz}
\end{figure}

One notable application is the use of spherical CNNs for analysis of data from the close-to-spherical Kamland-zen detector~\cite{KamLAND-Zen:2016pfg}.
This currently ongoing work incorporates a modification to best fit the detector geometry needs and has already demonstrated improvements in early stages~\cite{kamlandzen_spheres}. 
Given the nearly spherical shape of their detector, the authors of this work seek to correct a distortion created by the projection of the detector readout into a 2D pattern as seen in Figure~\ref{fig:kz}. 
They employ spherical CNNs for the task of signal-background classification. 
In a Spherical CNN, the kernel covers the entire phase-space by scanning in Euler angle rather than projecting the readout into 2D planes. 
Indeed, the use of spherical CNNs achieves background rejection of 71\% compared to 61\% for their original CNN~\cite{Li_2019_cnn}.

In some cases, other technologies already match the experimental needs substantially better than CNNs. 
This is the case for the networks used for analysis of IceCube Neutrino Observatory data, whose detector spatial sparsity and non-uniformity makes the data less than ideal for CNNs. 
Their deep learning application uses Graphic Neural Networks (GNNs) as a way to mitigate the effect of these features.
This is because GNNs are capable of dealing both with irregular geometry and graphs of different sizes, a feature which is seen in many of their events.
GNNs are designed to classify graphs, where the graph nodes define some element of the detector and the graph edges show some connection between elements~\cite{zhou2018graph}.
IceCube's GNN separates neutrino-induced  muons (their signal) from cosmic-ray shower-induced muons (their background), and compared the efficiency of the network to that of their standard reconstruction~\cite{choma2018graph}.
This GNN was able to identify 6.3 times more signal events and  provide a signal-to-noise rate 3 times larger.
A comparison with a CNN, which gave similar results to the traditional reconstruction, demonstrated that GNNs offer significant benefits in this application.

There are many examples of successfully overcoming this challenge of adapting  deep learning tools for neutrino data analysis. 
However, the approach taken for each application will continue to encounter different obstacles and considerations unique to the data and the tools chosen for analysis.
Careful consideration of these modifications continues to show substantial improvements to the direct application of image recognition technologies.  

\subsection*{Challenge 2 --- Quantifying Bias and Uncertainties}

One of the risks of applying machine learning is the possibility that the algorithms will learn information from the training data beyond what is intended. 
The risks associated with these techniques are neither new nor specific to the applications in our field. 
Furthermore, these challenges are starting to receive more attention in the broader community, industry applications, and government regulations around the world~\cite{mehrabi2019survey,usai,mlrace}.

The principal danger is that a dataset used for training contains information or implies underlying structure which may incorrectly bias the results, yet is learned by the network. 
This risk is true for all machine learning algorithms, but is particularly noteworthy with algorithms which perform feature extraction, such as CNNs.
While feature extraction is the main advantage of these algorithms, special attention is needed to mitigate this risk. 

Given that the features are abstract, their association with the physical traits of the data is largely unknown.
In addition, the networks used for detector data analysis are typically trained on simulated datasets of the events of interest. 
These simulations carry models and assumptions of the detector performance, the particle interactions, and other physical processes.
The challenge of quantifying and mitigating biases is particularly important to guarantee robust physical conclusions which are also model and generator independent. 

Apart from deliberate simulation choices, any effects introduced or distributions sculpted in the selection of the training data as well as any mismatch between simulated and real events coming from detection performance, particle interactions, calibration, or other effects have the possibility of propagating throughout the learning process undetected or unquantified. 
While some existing applications have devised tailored approaches at mitigating bias, 
standard and complete approaches to this important challenge are yet to be achieved. 

 
Minimizing known bias can be achieved through careful choices in the construction of input datasets. 
An example of this concept is the charge-only energy reconstruction CNN used by EXO-200~\cite{Delaquis_2018}.
This network is used to discriminate between Single-Site, and Multi-Site events and is found to outperform traditional reconstruction which had been used in previous publications~\cite{PhysRevLett.123.161802}. 
The network was initially trained using a simulated $^{228}$Th source. 
When a systematic study was performed with arbitrary resolution, disproportionately large improvements in resolution were found for events in the $^{208}$Tl peak with respect to other classes of events. 
In order to correct this, the network was instead trained on a calibration gamma ray source data, which acts as a proxy for various backgrounds, in the center of the detector. 
The CNN is tested on numerous samples, including simulation as well as $^{60}$Co, $^{208}$Tl, $^{226}$Ra and $^{228}$Tl calibration sources at a range of source locations. 
Having implemented this change to the training data, improved performance was found in the relevant energy range.


It is also possible that biases exist but are unknown to the developer, for instance, when an unintended artifact arises in the data. 
In contrast with the example above, where training sample composition was kept flat across different backgrounds, other characteristics might not be known to be skewed in unphysical ways. 
It is not possible to correct or quantify  bias which is not yet known, but techniques can be designed to either minimize them or look for them in the data. 

\begin{figure}
    \centering
    \includegraphics[width=1.0\textwidth]{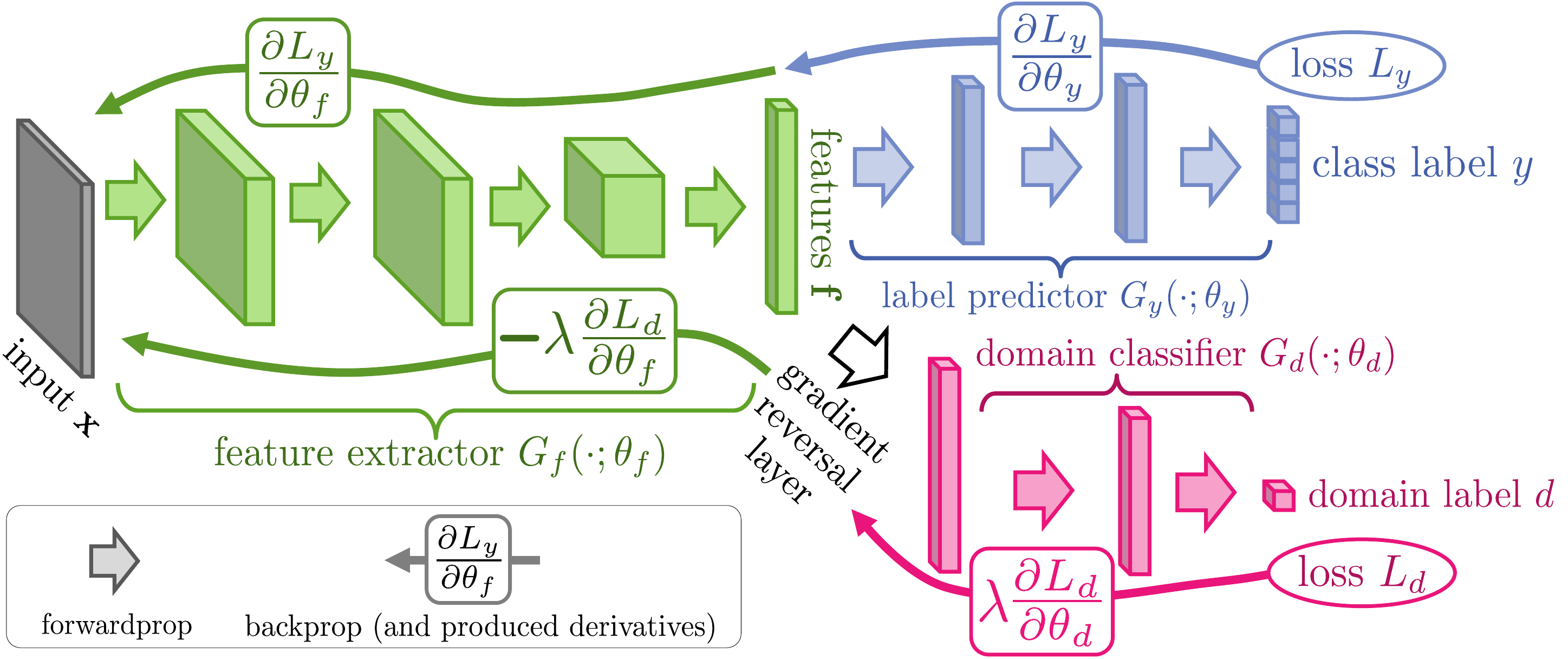}
    \caption{The basic structure of a domain adversarial neural network~\cite{ganin2015domainadversarial}.
    The feature extractor (green) and label predictor (blue) function just as a CNN.
    However, the domain classifier (pink) attempts to determine which of two domains the input is from.
    The gradient reversal layer discourages the network from classifying events using features that are unique to one of the two domains.}
    \label{fig:dann}
\end{figure}

The application of Domain Adversarial Neural Networks by the MINERvA experiment~\cite{Perdue_2018} is a prime example of unknown bias reduction. 
Here, the network is applied to the task of classification using a CNN, but a technique of bias reduction is employed.
The network is trained on both simulation and data.
The domain network, whose purpose is to distinguish between the data domain and simulation domain, is attached to the CNN.
It is expected to find features that result from errors or inconsistencies in the simulation. 
As shown in Figure~\ref{fig:dann}, the interplay between the two components discourages the task of classification to learn from any features that behave differently between the two domains. 

Another difficulty arises in designing tests that can reveal hidden biases learned by deep learning algorithms.
Ideally, this would look at the performance of the algorithm on data, but without a method of knowing the true nature of a data event, this is impossible (if such a method existed, we wouldn't need these algorithms in the first place!).
Instead, we must compare reconstructable quantities between data and simulations and look for signs of bias between the two, but which biases to look for is not obvious.
In addition, many experiments begin creating reconstruction algorithms before data taking has begun and others perform blind analyses where the algorithms must be optimized and validated without comparison to data.
While methods exist for constructing systematic uncertainties that address possible biases for many quantities, how to apply these methods to machine learning algorithms is not clear.

One example of a technique that uses both real and simulated data to search for bias is the muon-removed electron-added (MRE) technique used by the NOvA experiment~\cite{Sachdev:2015hpa}.
Two samples are created by overlaying simulations on real data events (MRE-on-data) and simulations on simulated events (MRE-on-simulation).
For each sample, an identified muon is removed from selected $\nu_\mu$ charged current events, leaving only the hadronic components of the interaction. 
The muon is substituted by a simulated electron of equivalent momentum overlaid on the events. 
Effectively, a comparison of the network performance between the MRE-on-data and MRE-on-simulation samples provides a measure of the bias-related effects introduced by data-simulation discrepancies in the hadronic component. 
The resulting difference in selection efficiency between the data overlay and the simulated overlay is less than 0.5\%.

Another data-based technique is to consider human-labeled datasets.
While we can't know the true identity of a real data event, a trained physicist can often identify with reasonable accuracy.
Comparisons of error patterns between humans and deep neural networks have shown differences between the two~\cite{humanbias}, which suggests differences between the unknown biases from humans to neural networks.
This is nevertheless a useful technique to search for large, unexpected biases in the outcome of the networks. 
The MicroBooNE experiment uses a liquid argon time projection chamber (LArTPC) detector for neutrino interactions~\cite{Fleming:2012gvl}.
They created a human-labeled dataset for validating a semantic segmentation network, a technique for classifying individual pixels in an image, trained on simulated neutrino interactions~\cite{Adams_2019}.
The disagreement between the performance of  the network and humans was less than 2\% in the misclassification of pixels.

\begin{figure}
    \centering
    \includegraphics[width=0.8\textwidth]{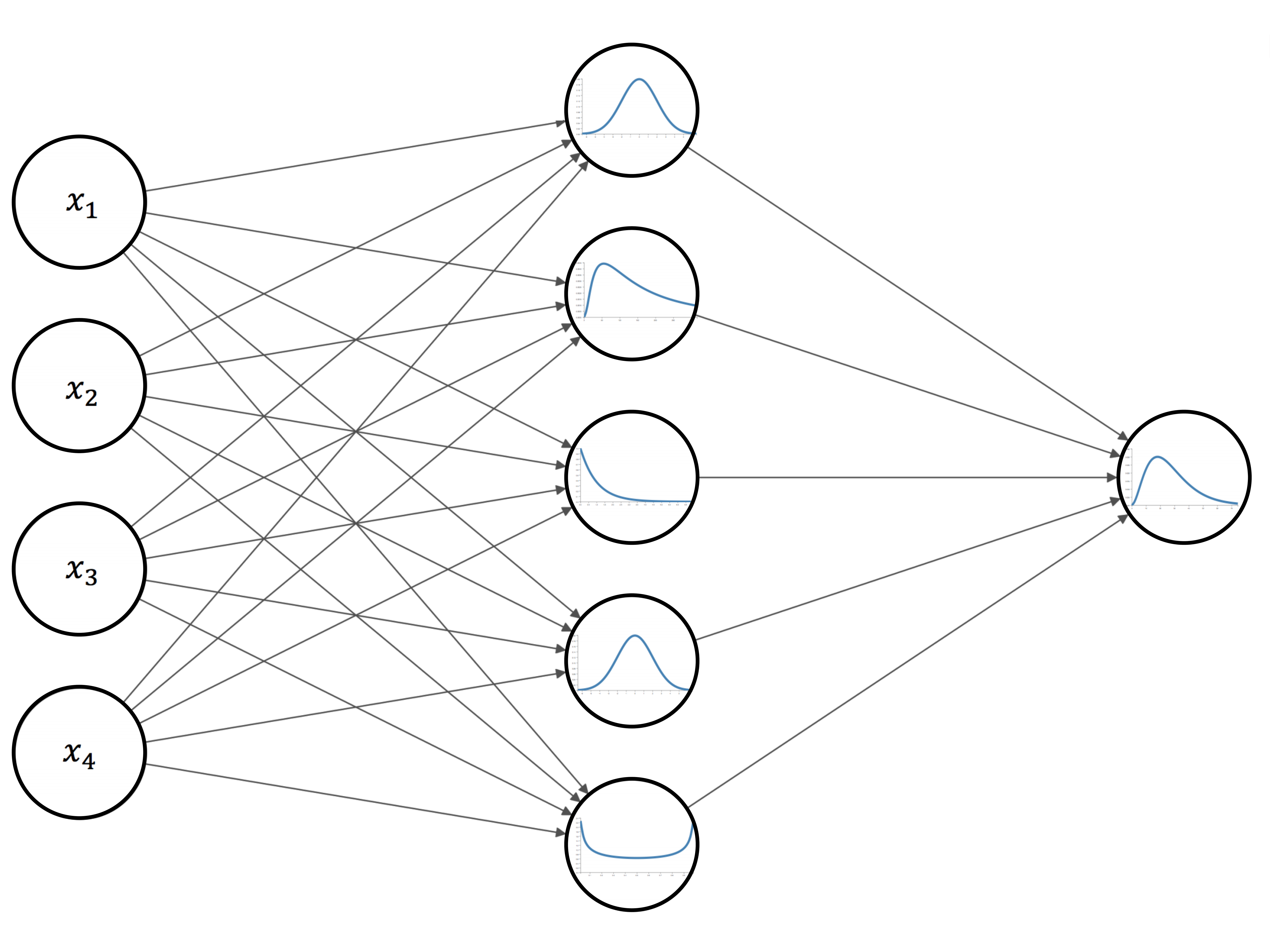}
    \caption{A depiction of a bayesian neural network. The fixed value weights in an artificial neural network are replaced by probability distribution functions. Thus, the output of each neuron is a probability distribution function with some most probable value and an uncertainty.}
    \label{fig:bnn}
\end{figure}

Finally, we consider uncertainties.
Quantifying the uncertainties associated with any measurement is an integral part of physics analyses.
Traditional neural networks, by design, output a single value. 
In some cases, they output high confidence scores on events that are well outside the phase space of samples they were trained on. 
While the output may be sensible in this case, it should incur a large uncertainty.
Bayesian neural networks are designed to address this concern~\cite{charnock2020bayesian}.
They replace the fixed value weights in the network with probability distribution functions, as shown in Figure~\ref{fig:bnn}.
The resulting output is, thus, also a probability distribution function which can be interpreted as a most probable value with some uncertainty.
This potential approach at including uncertainties has recently gained attention in the neutrino community~\cite{npml_bnn} with initial implementations currently being explored.

\subsection*{Challenge 3 --- Network Interpretability}

As machine learning models grow deeper, there is often a trade off between the performance of the algorithm and our ability to interpret its results.
Boosted decision trees, for example, are low-level machine learning models.
They can often inform the user of the relative importance of each input into the model, but may not have the accuracy that can be achieved with deeper models.
CNNs on the other hand, have achieved state of the art performance on many tasks, but the features extracted by the convolutional layers are abstract and challenging to interpret.
Some individual kernels can be connected to specific tasks, such as edge detection.
However, the features resulting from multiple convolutions are difficult to connect to topological characteristics or physical interpretations of the events.

This is particularly problematic in physics, where relating network features back to the underlying physics problem is important and sometimes necessary for a complete  understanding of the physical models. 
A better conceptual understanding of the physical features used by the network could tell us much about the physical processes which produced the features.
In addition, the understanding could aid to minimize or correct the inefficiencies in the performance of the algorithm.

\begin{figure}
    \centering
    \includegraphics[width=1.0\textwidth]{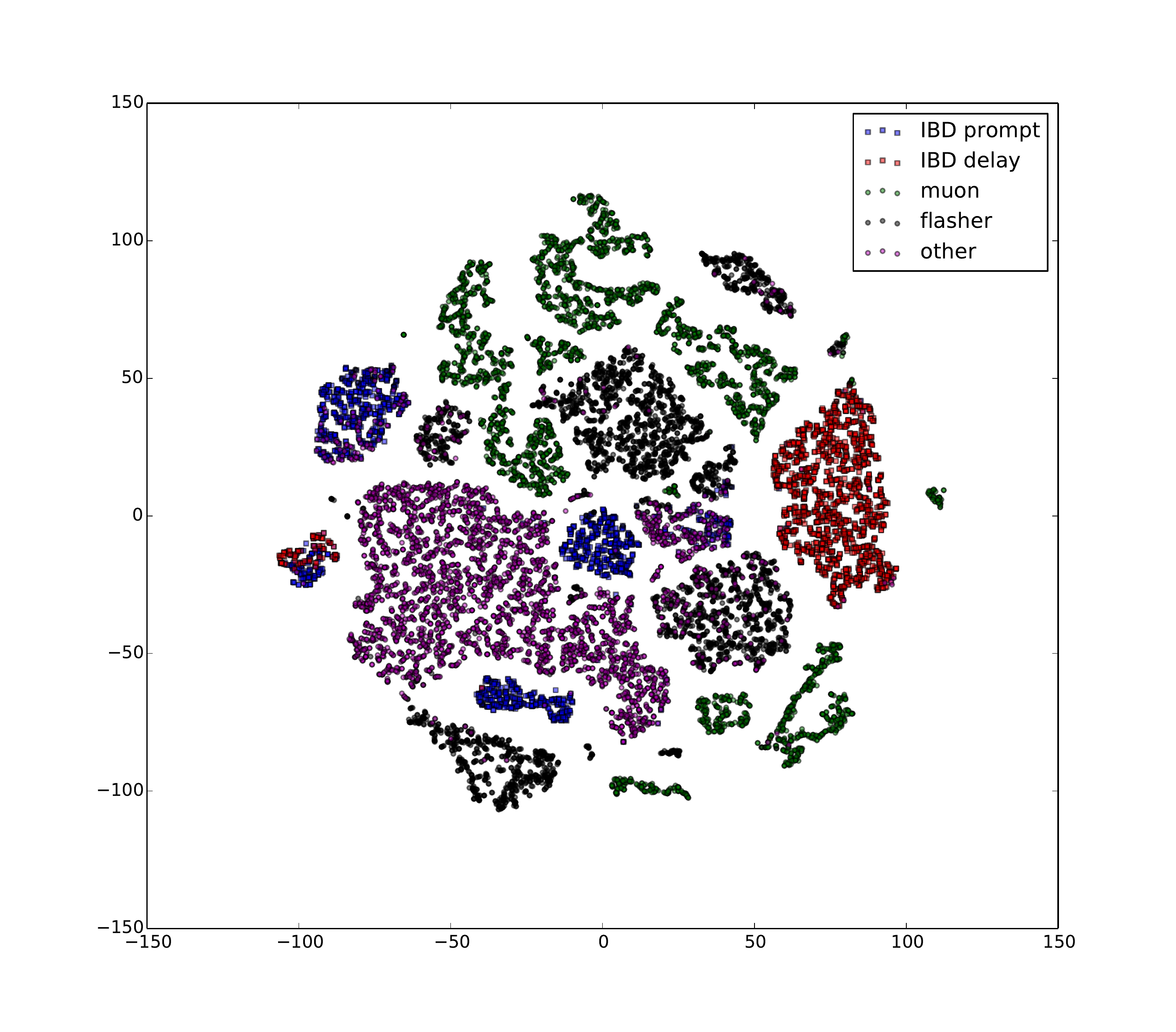}
    \caption{The t-SNE produced from the Daya Bay CNN to separate IBD signal from background noise. The t-SNE is a non-linear transformation used for dimensionality reduction. Visual separation in this space relates to separation in the high dimension feature space created by the network. Each point is labeled by it's true identity.}
    \label{fig:dayabay_tsne}
\end{figure}

A common method for interpreting the features extracted by the network is to perform dimensionality reduction.
The Daya Bay Reactor Neutrino Experiment is designed to detect anti-neutrinos produced by two nearby nuclear reactors~\cite{Cao_2016}.
They employ a CNN to separate inverse beta decay (IBD) events, the signal of interest, from noise within the detector~\cite{Racah:2016gnm}.
The features extracted by the network are transformed into two dimensions suing t-Distributed Stochastic Neighbor Embedding~\cite{tsne} (t-SNE).
The t-SNE method uses a non-linear transformation to reduce the dimensionality of data in a way that maintains the distance between points local to one another.
Figure~\ref{fig:dayabay_tsne} shows the result of this technique.
Class separation in this two dimension space, relates to topological information that the network has used to distinguish the different classes.

Another method of dimensionality reduction is Principal Component Analysis~\cite{Jolliffe1986} (PCA).
PCA is a linear change of basis where the new basis has vectors along the dimensions of maximum variation in the data.
Often only a few of the basis vectors are needed to explain most the variation in the data.
Critically, the new basis vectors are orthogonal meaning each has a unique contribution to the variation in the data.
PCA is often performed on the input data to a network to reduce the number of inputs needed to a smaller set of independent values which are most important to the task.
PCA can also be performed on the network extracted features to reduce the dimensionality for visualizations in a similar way to t-SNE.

In addition, some qualitative methods try to determine which features of the input are most relevant to the output.
This is particularly important for CNNs doing image recognition where determining which topological features of the input are most important to the network output.
One method of doing this is to occlude regions of the input image and determine how the various output scores change.
This technique is often called an occlusion test.
Another technique is to use the network itself to determine these valuable features.
Salience maps~\cite{simonyan2013deep} determine the gradient of the output score from the network with respect to each of the input pixels.
These maps can show where the network is "looking" to construct it's features.
Interestingly, these sometimes show that CNNs do not look at the primary object in an image, but instead at the surrounding context.
If some objects are commonly found in the same context, then the context can be used as the primary discriminator to classify that object.

\subsection*{Challenge 4 --- Computational \& System Constraints} 

As mentioned in section~\ref{sec:new-ml}, the latest developments in deep learning are largely driven by improvements in GPU technology where the many computations needed for large networks can be done in parallel.
Deep neural networks often perform $\mathcal{O}\left(10^9\right)$ floating point operations.
This is compounded by the amount of data collected in particle physics experiments.
Modern neutrino experiments record billions of events which require evaluation by various reconstruction and analysis algorithms.
Many experiments perform these evaluations on large-scale computing grids on CPUs.

While neural networks have expanded the capabilities of many neutrino experiments, this computing limitation provides a bottleneck to widespread use of very deep neural networks.
Here we consider three methods to alleviate this concern.

One potential solution is to expand the availability of GPUs.
Small GPU clusters used for training neural networks are becoming more common.
However, these are not enough to match the production needs of many experiments.
Larger availability of GPU clusters would enhance the ability of experiments to utilize large neural network based algorithms.

Another possibility is to enhance the physics output from these algorithms.
As discussed throughout this manuscript, machine learning based methods often show significant improvements over traditional methods.
One way to improve performance is to maximize the primary task algorithm, but the implementation of multi-task algorithms could be a promising way to enhance the total physics output from an individual algorithm.
The Deep Underground Neutrino Experiment (DUNE) is a future neutrino oscillation experiment currently in R\&D stages~\cite{Abi:2020wmh} for its LArTPC detector.
The DUNE experiment employs a CNN for identification of neutrino interaction flavor in their detector~\cite{collaboration2020neutrino}, which achieves more than 85\% efficiency of $\nu_\mathrm{e}$ charged-current events in the energy range of interest.
In addition to flavor classification, the algorithm also outputs the sign of the neutrino, the type of interaction, and the amounts of each particle in the final state.
In total, the network has seven outputs at very little additional computational cost since each output uses the same set of features extracted by the network.

\begin{figure}
    \centering
    \includegraphics[width=0.6\textwidth]{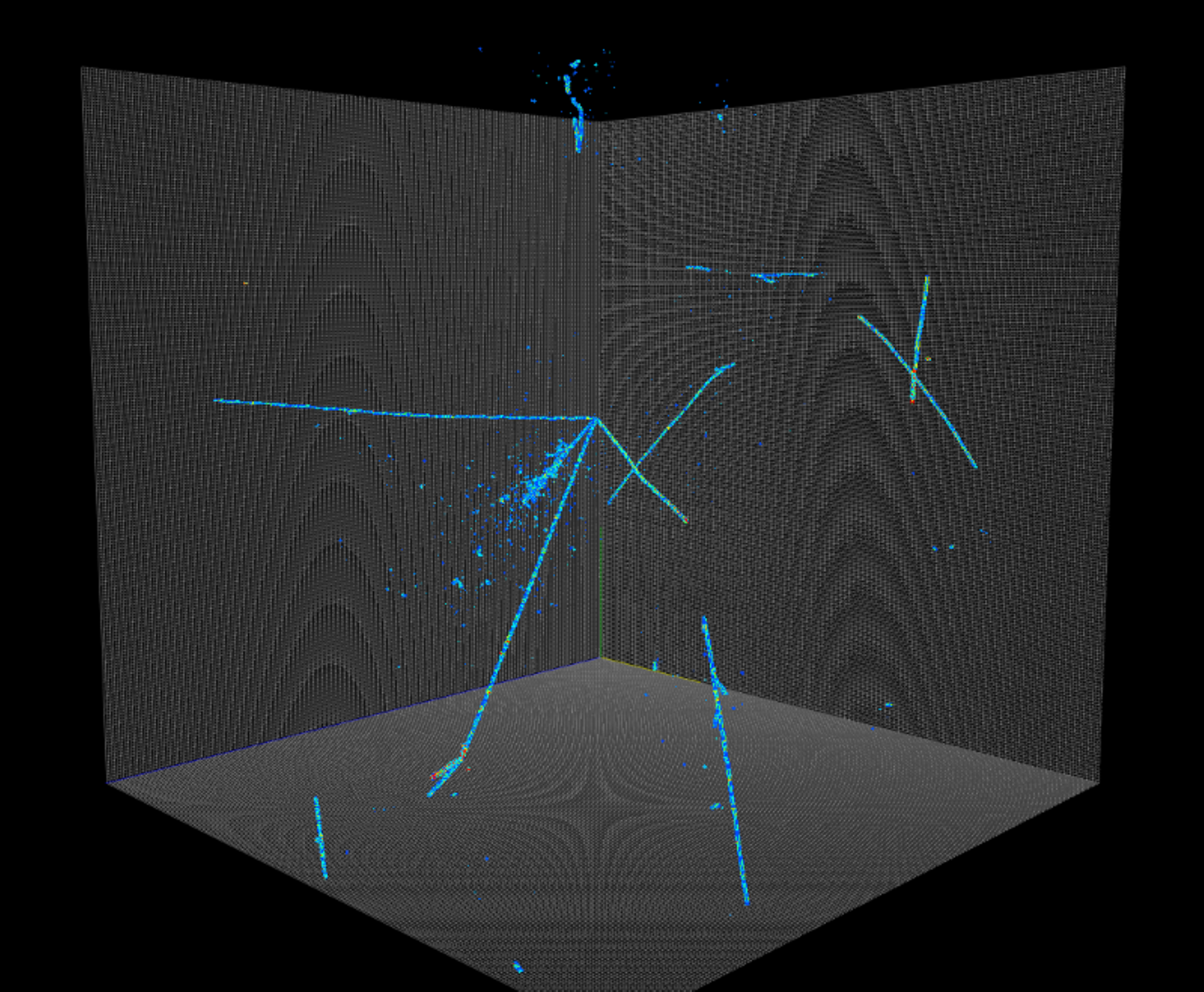}
    \caption{An example neutrino event simulated in a liquid argon time projection chamber.
    Shown are energy deposits from charged particles traversing the detector medium.
    Less than 1\% of the detector is active in this event.}
    \label{fig:uboone_event}
\end{figure}

Reducing the computational cost of the algorithms is another option which would reduce the total computational need of experiments.
Using smaller networks is one option, but this comes at the cost of performance.
Instead, considerations can be given to the type of data acquired by experiments.
LArTPC detectors, such as those used by DUNE or the Short Baseline Neutrino program~\cite{Machado_2019}, have very low occupancy, the fraction of active detector readout from an event.
These events are globally sparse, $<1\%$ sparsity, but locally dense, in the region of the detector where the event occurred.
An example of an event recorded in a LArTPC is in Figure~\ref{fig:uboone_event}.
This means that typical CNNs will waste much computation time multiplying or summing together zeros.
It's been shown that using submanifold sparse convolutional networks~\cite{Domin__2020} can reduce the inference time of these networks by a factor of 30 and the memory cost by more than 300.
These sparse convolutional networks are designed for use with sparse data and only perform convolution operations in regions with activity.

Finally, we consider the use of open datasets in algorithm development.
Open datasets are commonly used to benchmark algorithm performance in data science applications~\cite{imagenet_cvpr09}.
Despite increasing efforts from a handful of experiments to provide such datasets for analysis~\cite{dlp_opendata}, there are still many restrictions surrounding data-sharing in the field.
The lack of available data sets negatively impacts the ability of researchers to develop and publish improved machine learning techniques specific to particle physics applications and significantly hinders progress in developments requiring real data, such as bias assessment.
Open data sets would not only enable these advancements to be developed further, but it would significantly encourage beneficial multi-disciplinary collaboration which would surely improve the quality of physics of our our experiments.

\section{Opportunities going forward}

The use of machine learning and currently deep learning algorithms for neutrino experiment data analysis is on the rise. 
We have presented an overview of the impacts of these techniques in the field through a description of the challenges and opportunities associated with their usage. 

\subsection*{Opportunity 1 --- Impact to Physics and Technology}


The application of machine learning tools to neutrino physics is also relevant to the process of experiment design and proposal, which brings about opportunities to further impact the capabilities of future experiments.  
The next generation of neutrino experiments will introduce needs and challenges beyond what the field has encountered. 
Massive detectors designed to measure neutrino oscillations will redefine the challenges of data rates and data management and will continue to look for ways to expand their physics program~\cite{dunetdr,hyperktdr}. 
Neutrino-less double beta decay experiments at and beyond the ton-scale will require exceptional rejection of radioactive backgrounds beyond what has ever been achieved. 
The emerging field of multi-messenger astronomy will further encourage experiments to expand their sensitivity to signals beyond their current reach.

Much like previous generations, this generation of experiments will only be possible by pushing technological frontiers. 
This presents opportunities for the field of particle physics and machine learning, which could cement the synergy between the two fields in mutually beneficial ways. 

\begin{figure}
    \centering
    \includegraphics[width=0.49\textwidth]{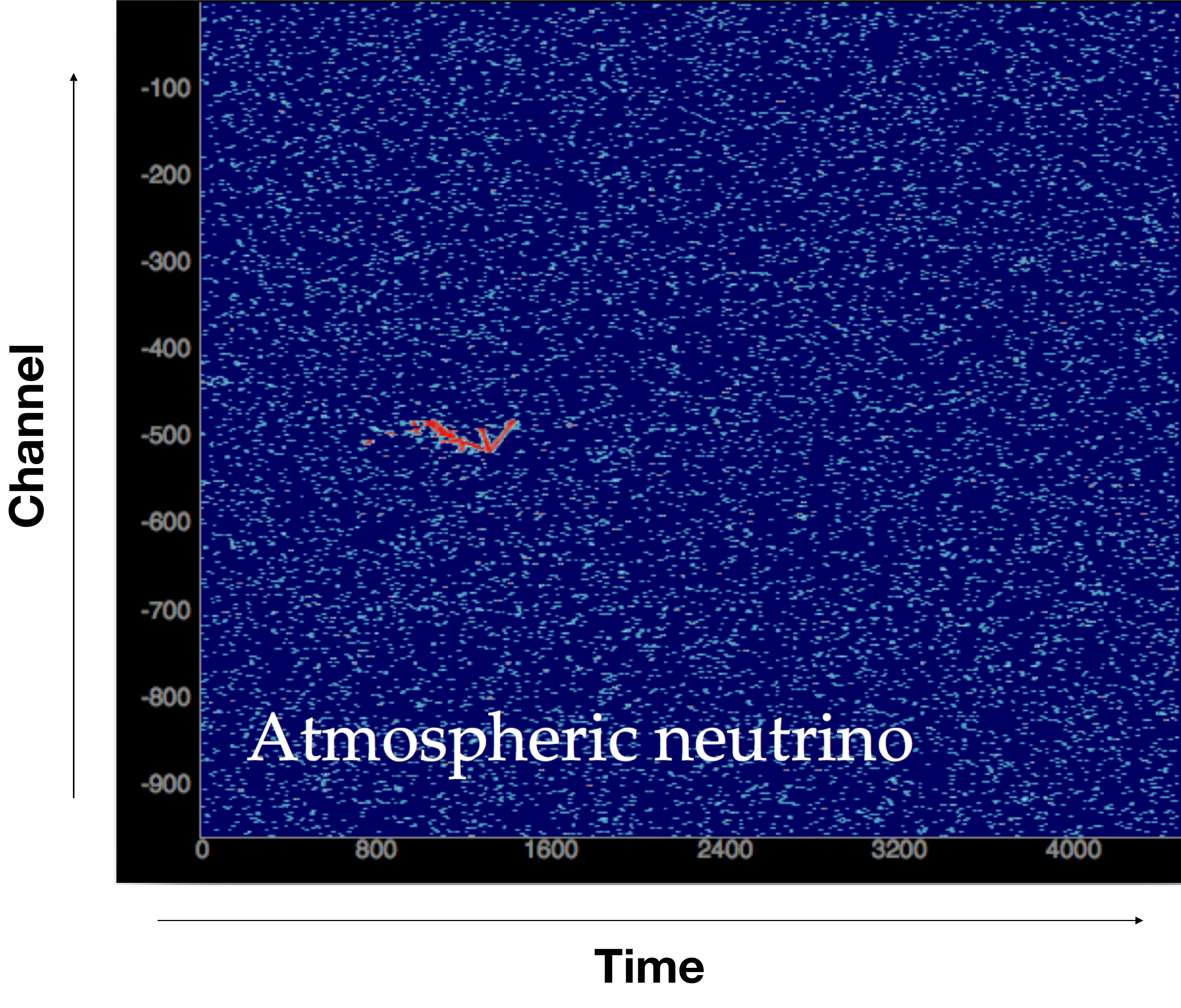}
    \hfill
    \includegraphics[width=0.49\textwidth]{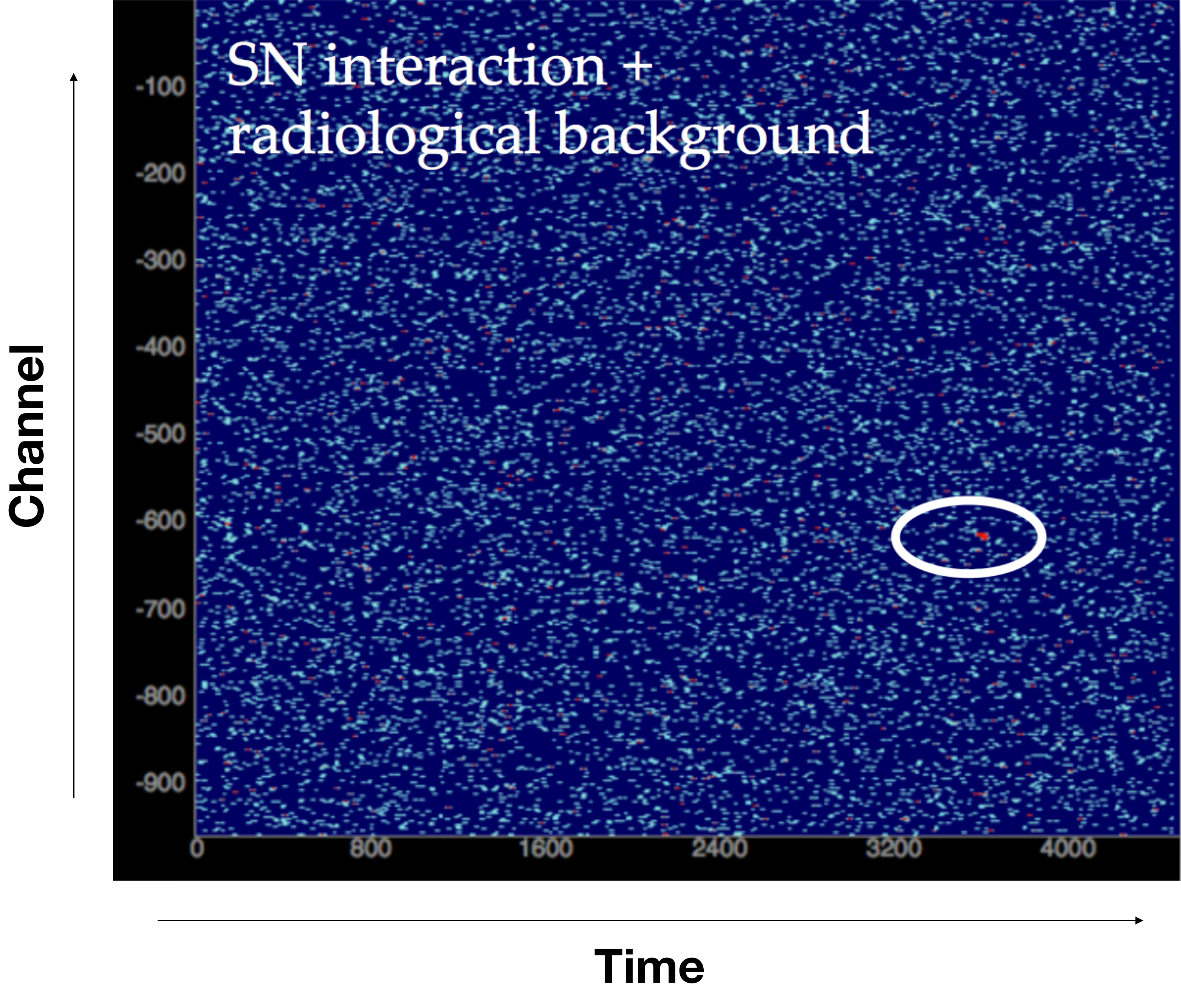}
    \caption{Left: A high energy atmospheric neutrino interaction in the DUNE LArTPC. Right: A low energy supernova neutrino interaction in the DUNE LArTPC.}
    \label{fig:dune_sn}
\end{figure}

An interesting example of research and development (R\&D) involving machine learning is their application on the hardware trigger being developed for the DUNE experiment. The large data rates expected on DUNE detectors currently constrain the energy range available for analysis. 
Figure~\ref{fig:dune_sn} shows a single DUNE data frame. 
The majority of the electronics noise as well as radioactive backgrounds are safely below the energies of the accelerator neutrinos DUNE is designed to study. 
However, there are also interesting signals on the MeV-scale energy range which could potentially be studied such as supernova neutrinos, solar neutrinos~\cite{Capozzi_2019}, and neutrino-less double beta decay~\cite{dunebeta}. 
Unfortunately, it is possible that the currently available hardware for data acquisition systems will require the elimination of much of the low energy noise from DUNE's data stream at the trigger level in order to maintain manageable data rates. 
However, if the physics reach of DUNE could be extended to study low energy signals, it could produce world leading measurements of solar neutrino oscillations. 

In order to enable a DUNE low energy program, data acquisition hardware will need to sustain high rates at low down times. 
Research into the applications of deep learning to hardware triggers and data acquisition for DUNE is ongoing to resolve this issue. 
Hardware acceleration as well as well as optimal implementation of deep learning algorithms on FPGAs and GPUs is being explored. 
This work is exploring the possibility of online data analysis capable of process up to tens of terabits per second, aided by the capabilities of CNNs to tackle high rate image processing~\cite{dunetrigger}.

The capabilities of this trigger may well define whether low energy signals will be available to explore on DUNE. 
Thus, the usage of machine learning algorithms might significantly contribute to not just improving performance of existing analyses, but expanding the physics program that is available to experiments. 
While there is community consensus on some of the challenges machine learning will need to address going forward, we are only starting to recognize that machine learning development is an integral part of neutrino physics research~\cite{Albertsson:2018maf}.
The continued active pursuit of R\&D involving machine learning applications might significantly change the neutrino physics landscape in the coming decades.

\subsection*{Opportunity 2 --- What Physics Can Contribute to Machine Learning}

The unique nature of the problem set and analysis strategies of neutrino physics (and particle physics) experiments brings about the potential to contribute new knowledge and applications to the field of computer science. 
Two aspects drive this opportunity:

1. Quantitative results and careful statistical analysis. 
Statistical precision is one of the hallmarks of particle physics experiments. Carefully quantifying results and uncertainties becomes even more important as neutrino experiments move into the precision era. 
As we develop tools and techniques to address the challenge of bias assessment and uncertainty quantification for our needs, these developments will surely inform the broader picture of secure, ethical, and responsible treatment of machine learning beyond scientific applications~\cite{aieithics,algethics}.

2. Customizable simulated datasets corresponding to real physical data. 
The majority of industry applications of machine learning are developed, tested, and applied in real-world datasets. 
Training usually employs labeled data of the same type as that to which network will be applied. 
In contrast, neutrino experiments usually construct and train most of their analysis infrastructure on simulated data that resembles the expected data. 
The detail to which these simulations are tunable is especially relevant to the study of machine learning algorithms. It provides the opportunity to study their behavior under controlled modifications in the training samples, which could greatly contribute to the challenge of explainability in and outside the field. 

\subsection*{Opportunity 3 --- Innovations}


There are also opportunities in the area of overlap between the problem sets of neutrino physics and machine learning. 
It is no surprise that we are starting to develop machine learning inspired tools which can be applicable outside neutrino physics. 

\begin{figure}
    \centering
    \includegraphics[width=1.0\textwidth]{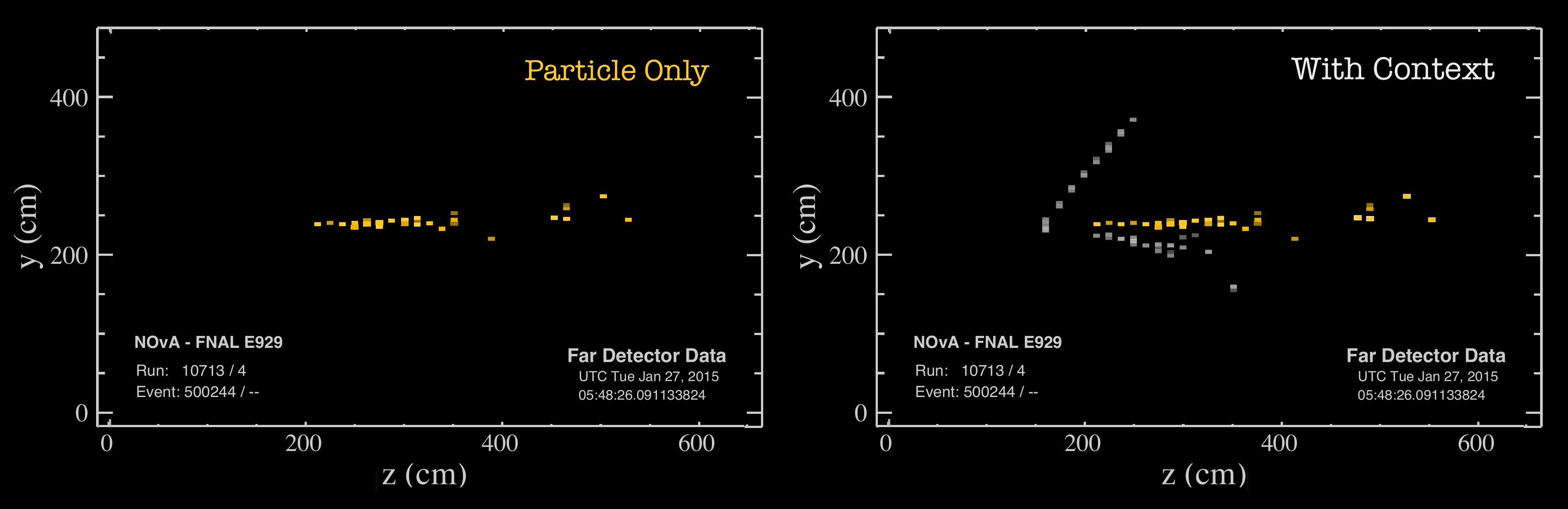}
    \caption{One view of a neutral current event with a $\pi^0$ decay in the NOvA detector. Left: Just one of the particles produced by the decay. Right: The same particle with the context from the rest of the event (grey) shown. The knowledge from the context, such as the particle separation from the vertex, make it clear that this is a photon shower.}
    \label{fig:nova_context}
\end{figure}

For example, applications of machine learning to NOvA detector data have been further explored, specifically targeting single particle identification within clusters of particles. 
As shown in Figure~\ref{fig:nova_context}, each cluster of energy depositions in an interaction needs to be further analyzed to identify its producer. 
In this case, knowledge of the single particle cluster is useful, but there is much to be gained from providing some context to the classification network. 
In a recent publication~\cite{Psihas_2019}, the authors demonstrate a technique to add context information to a CNN input and how to implement the Siamese concept to take advantage of ``particle-only" as well as ``context view" of the inputs. 
This technique is the first to employ a Siamese architecture for the addition of context. 
As such, it is a contribution to both fields. 
In the neutrino physics application, it was found that adding context to the inputs improved the identification efficiency of particles by up to 11\%.

Opportunities for the synergy between neutrino physics and machine learning are plentiful. 
The deeper appreciation for the complexity and overlap of each of their problem sets may continue to give way to enhanced advances for both fields.

\section{Acknowledgements}
The authors thank Justin Vasel for his helpful review of this manuscript. The authors would like to acknowledge Georgia Karagiorgi for productive conversations about the exciting directions of R\&D for accelerated hardware as well as Taritree Wongjirad, Kazuhiro Terao, and Tingjun Yang for their useful insight of the challenges of LArTPC applications. 
Fermilab is operated by Fermi Research Alliance, LLC under Contract No. DE-AC02-07CH11359 with the U.S. Department of Energy, Office of Science, Office of High Energy Physics. The United States Government retains and the publisher, by accepting the article for publication, acknowledges that the United States Government retains a non-exclusive, paid-up, irrevocable, world-wide license to publish or reproduce the published form of this manuscript, or allow others to do so, for United States Government purposes.

\bibliographystyle{unsrt}
\bibliography{review}

\end{document}